\documentclass[structabstract]{aa}

\usepackage{graphicx}
\usepackage{txfonts}
\begin{document}
   \title{Effect of NLTE model atmospheres on photometric amplitudes\\ and phases of early B-type pulsating stars}

   \author{J. Daszy\'nska-Daszkiewicz
          \and
          W. Szewczuk
          }

   \institute{Instytut Astronomiczny, Uniwersytet Wroc{\l}awski,
              ul. Kopernika 11, 51-622 Wroc{\l}aw, Poland\\
              \email{daszynska@astro.uni.wroc.pl, szewczuk@astro.uni.wroc..pl
               }
             }

   \date{Received ...; accepted ...}


  \abstract
   {Amplitudes and phases of the light variation of a pulsating star in various photometric passbands
contain information about geometry of observed modes. Because oscillation spectra of early B-type main sequence stars
do not exhibit regular patterns, these observables are very often the only ones from which mode identification can be derived.
Moreover, these data can yield valuable constraints on mean stellar parameters, subphotospheric convection,
microphysics and atmospheres.}
{We study all possible sources of inaccuracy in theoretical values of
the photometric observables, i.e. amplitude ratios and phase differences, of early B-type main sequence pulsators.
Here, we discuss effects of parameters coming from both models of stellar atmospheres and
linear nonadiabatic theory of stellar pulsation. In particular, we evaluate for the first time
the effect of the departure from the LTE approximation.}
   {The photometric amplitudes and phases are calculated from the semi-analytical formula for the light variation.
   The atmospheric input comes from line-blanketed, LTE and NLTE plane-parallel, hydrostatic models.
   To compute the limb-darkening coefficients for NLTE models, we use the Least-Square Method
   taking into account the accuracy of the flux conservation.
   The linear nonadiabatic stellar pulsations were computed by means of Dziembowski code.
   We consider the OPAL and OP opacity tables and two determinations of the solar chemical mixture: GN93 and AGSS09.}
   {We present effects of NLTE atmospheres, chemical composition and opacities on theoretical values
   of the photometric observables of early B-type pulsators. To this end, we compute tables
   with the passband fluxes, flux derivatives over effective temperature and gravity as well as the non-linear limb-darkening coefficients
   in 12 most often used passbands, i.e. in the Str\"omgern system, $uvby$, and in the Johnson-Cousins-Glass system, $UBVRIJHK$.
   We make these tables public available at the Wroc{\l}aw HELAS Web page \thanks{http://helas.astro.uni.wroc.pl}.}
   {In the case of radial modes, effects of uncertainties in model atmospheres
   are much smaller than those coming from the pulsation theory. In turn, for nonradial modes effects of NLTE
   become more important and they are most significant for dipole modes.
   Therefore, if one wants to construct an accurate seismic model and gain more constraints,
   all inaccuracies in the photometric observables have to be allowed for.}


\titlerunning{Effect of NLTE on photometric amplitudes and phases of early B-type pulsators}
\maketitle
%

\section{Introduction}
Data on the photometric and spectroscopic variations of a pulsating star bring information on
frequencies of excited modes and their geometry. The latter property is particularly important
if oscillation spectra are sparse and lack equidistant patterns. This is the case for main
sequence pulsators of early B spectral type, i.e. $\beta$ Cephei variables.
Additionally, using data on the photometric amplitudes and phases one can constrain
mean stellar parameters, test efficiency of convection transport in the case of cooler pulsators, like
$\delta$ Scuti stars, (Daszy\'nska-Daszkiewicz, Dziembowski, Pamyatnykh 2003) or test opacity data
in the case of B-type pulsators (Daszy\'nska-Daszkiewicz, Dziembowski, Pamyatnykh 2005).

Ones of the most popular tools to identify a pulsation mode
are the amplitude ratios and phase differences in various photometric passbands.
In the zero-rotation approximation, these observables depend on the mode degree, $\ell$,
but are independent of the azimuthal order, $m$, the intrinsic mode amplitude, $\varepsilon$,
and the inclination angle, $i$.

The semi-analytical expression for the bolometric light variation was formulated
by Dziembowski (\cite {dziemb}). Then, Balona \& Stobie (\cite{balst})
and Stamford \& Watson (\cite{stwa}) expanded it for the light
variation in the photometric passbands. They showed that modes with
different values of $\ell$ are located in separated parts on the amplitude ratio
$vs.$ phase difference diagrams.
Subsequently, this method has been applied to various types of pulsating stars
by Watson (\cite{watson}). Cugier, Dziembowski \& Pamyatnykh (\cite{cdp})
improved the method by including nonadiabatic effects
in calculations for the $\beta$ Cephei stars.
Effects of rotation on photometric observables were studied by Daszy\'nska-Daszkiewicz et al. (\cite{dd2002})
for close frequency modes and by Townsend (\cite{townsend2003})
and Daszy\'nska-Daszkiewicz, Dziembowski \& Pamyatnykh (\cite{dd2007})
for long-period g-modes.
To compute values of the photometric amplitudes and phases we need two kinds of input.
The first one comes from stellar model atmospheres and the second one from computations
of stellar pulsation. Both data contain various sources of uncertainties.

The goal of this paper is to examine for the first time an influence of NLTE effects on theoretical values
of the photometric amplitude ratios and phase differences for early B-type main sequence pulsators.
We check also effects of metallicity and microturbulent velocity in the atmosphere.
Subsequently, we compare these effects with uncertainties coming from the linear nonadiabatic theory
of stellar pulsation. As an example, we consider main sequence models with a mass of 10 $M_\odot$
and low degree modes with $\ell$=0, 1, 2. Here, we neglect all effects of rotation on pulsation.

The structure of the paper is the following.
In Section\,2, we recall the linear formula for the pulsation complex amplitude in a photometric band.
Section\,3 contains description of NLTE model atmospheres and results of our computations
of the band fluxes, corresponding flux derivatives over effective temperature and gravity,
and the nonlinear limb darkening coefficients, in 12 photometric passbands: $uvbyUBVRIJHK$.
Tables with these data can be downloaded from the Wroc{\l}aw HELAS web page.
Moreover, we study effects of temperature, gravity, NLTE, atmospheric metallicity and
microturbulent velocity on the above mentioned quantities.
How these atmospheric uncertainties translate into the pulsation photometric observables,
i.e. amplitude ratios and phase differences, is discussed in Section\,4.
Inaccuracies in the photometric observables connected with the linear non\-adiabatic
theory of stellar pulsation are presented in Section\,5.
We end with conclusions in Section\,6.

\section{Light variation due to stellar pulsation}
Stellar pulsations cause changes of temperature, normal to the surface element
and pressure. If pulsations are linear and all effects of rotation on pulsation can be ignored,
then the total amplitude of the light variation in the passband $x$
can be written in the following complex form (Daszy\'nska-Daszkiewicz et al. \cite{dd2002}):
\begin{equation}
  \label{eq1}
  \mathcal{A}_x(i)=-1.086 \varepsilon Y_{\ell}^m(i,0)b_{\ell}^x(D_{1,\ell}^x f+D_{2,\ell}+D_{3,\ell}^x),
\end{equation}
where $\varepsilon$ is the intrinsic mode amplitude, $Y_{\ell}^m$ -- the spherical harmonic
and $i$ -- the inclination angle.
The amplitude is given by $abs(\mathcal{A}_x)$ and phase by $arg(\mathcal{A}_x)$.
The $D_{1,\ell}^x\cdot f$ product stands for temperature changes, where
\begin{equation}
  \label{eq2}
  D_{1,\ell}^x=\frac{1}{4} \frac{\partial \log(\mathcal{F}_x|b_{\ell}^x|)}
  {\partial \log T_{\rm eff}}.
\end{equation}
${\cal F}_x$ is the flux in the passband $x$ and $f$ is the nonadiabatic complex parameter
describing the amplitude of the radiative flux perturbation to the radial displacement
at the photosphere level
\begin{equation}
\label{f_par}
\frac{ \delta {\cal F}_{\rm bol} } { {\cal F}_{\rm bol} }=
{\rm Re}\{ \varepsilon f Y_\ell^m(\theta,\varphi) {\rm e}^{-{\rm i}
\omega t} \}.
\end{equation}
Geometrical term, $D_{2,\ell}$, is given by
\begin{equation}
  \label{eq3}
  D_{2,\ell}=(2+\ell)(1-\ell),
\end{equation}
and the pressure term, $D_{3,\ell}^x$, by
\begin{equation}
  \label{eq4}
  D_{3,\ell}^x=-\left( 2+ \frac{\omega ^2R^3}{GM} \right) \frac{\partial \log(\mathcal{F}_x|b_{\ell}^x|)}
  {\partial \log g}.
\end{equation}
$b_{\ell}^x$ is the disc averaging factor defined by
\begin{equation}
  \label{eq5}
  b_{\ell}^x= \int _0^1 h_x (\mu) \mu P_\ell(\mu) d\mu
\end{equation}
where $h_x(\mu)$ is the limb darkening law, $P_\ell$ is the Legendre polynomial
and $\mu$ is a cosine of the angle between a line of sight and the emergent intensity.
Remaining parameters have their usual meaning.

In the above expressions, we can distinguish two sorts of input parameters
needed to compute theoretical values of the photometric amplitudes and phases.
The first input is derived from models of stellar atmospheres and these
are the flux derivatives over effective temperature and gravity (Eq.\,\ref{eq2} and \ref{eq4}),
as well as limb-darkening and its derivatives (Eq.\,\ref{eq2}, \ref{eq4} and \ref{eq5}).
The second input comes from the nonadiabatic theory of stellar pulsation
and this is the $f$-parameter (Eq.\,\ref{eq1} and \ref{f_par}).

In this paper, we used the Warsaw-New Jersey evolutionary code
and the linear nonadiabatic pulsation code of Dziembowski (1977).
We considered \linebreak[4] opacity tables from OPAL (Iglesias \& Rogers 1996) and OP (Seaton 2005) projects,
and two determinations of the solar chemical composition: GN93 by Grevesse \& Noels 1993
and AGSS09, a recent one by Asplund et al. 2009.
As for models of stellar atmospheres, we considered Kurucz models (Kurucz \cite{kurucz})
and TLUSTY models (Lanz \& Hubeny \cite{lanzhubeny}).

\section{NLTE line-blanketed model atmospheres}
The most widely used models of stellar atmospheres are line blanketed, plane-parallel, hydrostatic models
of Kurucz (2004) computed within an approximation of local thermodynamic equilibrium (LTE).
However, in the case of atmospheres of early B-type stars, effects of the departure from LTE
and a proper treatment of line opacity become important.
Escalation of the quality of spectro-photometric observations calls for
a need of high resolution and accurate models of stellar atmospheres.

A grid of non-LTE (NLTE) model atmospheres were computed by Lanz \& Hubeny (\cite{lanzhubenyO})
and Lanz \& Hubeny (\cite{lanzhubeny}).
These are metal line-blanketed, plane-parallel, hydrostatic model atmospheres of O-type stars (OSTAR2002)
and of early B-type stars (BSTAR2006), respectively. In their computations, they adopted a solar chemical mixture
by Grevesse \& Sauval ({\cite{grevese}), helium to hydrogen abundance of He/H=0.1 by number
and two values of the microturbulent velocity, $\xi_t$=2 and 10 km/s.
Lanz \& Hubeny (\cite{lanzhubenyO}, \cite{lanzhubeny}) showed that in the case of OB stars,
a neglect of NLTE effects causes differences not only in spectral lines but also in the continuum level.
In the near ultraviolet (the Balmer continuum), the LTE fluxes are up to $10\%$ higher than the
NLTE counterparts. In turn, in the far and extreme ultraviolet (the Lyman continuum), the LTE fluxes are lower
than the NLTE ones. For more details see Lanz \& Hubeny (\cite{lanzhubenyO}, \cite{lanzhubeny}).

The grid of the OSTAR2002 models were computed for 12 values of effective temperatures appropriate to O-type stars,
i.e. between 27500 and 55000 K with a step of 2500 K, 8 values of surface gravities from $\log g$=3.0 to 4.75
with a step of 0.25 dex and one microturbulent velocity, $\xi_t$=10 km/s.
Moreover, 10 values of the atmospheric metallicity were considered, $(Z/Z_{\odot})_{\rm atm}$=2, 1, 0.5, 0.2, 0.1, 1/30, 1/50, 1/100, 1/1000, 0.0,
where $Z$ is the metal abundance by mass and $Z_{\odot}$ is the solar value.

The grid of the BSTAR2006 models contains 16 values of effective temperatures between 15000 and 30000 K
and a step of 1000 K, 13 values of surface gravities in the range from $\log g=1.75$ to 4.75 dex and a 0.25 dex step.
In this case, 6 values of metallicity were considered, $(Z/Z_{\odot})_{\rm atm}$=2, 1, 0.5, 0.2, 0.1, 0.0, and two values
of the microturbulent velocities, $\xi_t$=2 km/s and $\xi_t$=10 km/s.
Models with $\xi_t$=10 km/s were computed for two sets of chemical mixture but only for B-type supergiants ($\log g \le 3.0$).
The first mixture was the same as for $\xi_t$=2 km/s, i.e. the solar composition,
and the second one was enriched in helium and nitrogen, and depleted in carbon.

The lower limit for gravity at a given effective temperature was determined approximately by the Eddington limit.
The TLUSTY code used by Hubeny and Lanz (1995) becomes unstable near this limit.

\subsection{The passband fluxes and their derivatives}
We computed the fluxes, ${\cal F}_x$, for the BSTAR2006 NLTE models in 12 commonly used photometric passbands,
i.e. in the Str\"omgren ($uvby$) and Johnson-Cousins-Glass ($UBVRIJHK$) systems, according to the formula
\begin{equation}
\label{integ}
{\cal F}_x=\frac{\int\limits^{\lambda_1}_{\lambda_2}{\cal F}(\lambda)S(\lambda)\,d\lambda}{\int\limits^{\lambda_1}_{\lambda_2}S(\lambda)\,d\lambda},
\end{equation}
where $S(\lambda)$ is the response function of the passband $x$, adopted from
the Asiago Database on Photometric Systems (Moro \& Munari \cite{moro}).
The integral in Eq.\,(\ref{integ}) is computed in the wavelength range from $\lambda_1$ to $\lambda_2$ where
$S(\lambda)$ has non-zero values.

We considered the whole grid of effective temperature, $T_{\rm eff}$, gravity, $\log g$,
metallicity, $(Z/Z_{\odot})_{\rm atm}$, and microturbulent velocity, $\xi_t$, for the BSTAR2006 models.
Table\,\ref{strumienie} summarizes our results.
The file names are coded by values of $(Z/Z_{\odot})_{\rm atm}$ and $\xi_t$.
We have adopted the same designations as Lanz \&Hubeny (\cite{lanzhubeny}), i.e.
BC, BG, BL, BS, BT, BZ denote $(Z/Z_{\odot})_{\rm atm}$=2, 1, 0.5, 0.2, 0.1 and 0, respectively,
whereas v2 and v10 correspond to $\xi_t= 2$ km/s  and 10 km/s, respectively.
The index CN marks a model enriched in helium and nitrogen, and depleted in carbon.
Each file contains the following columns: line number, effective temperature, $T_{\rm eff}$,
logarithm of the surface gravity, $\log g$, metallicity, $(Z/Z_{\odot})_{\rm atm}$, microturbulent velocity, $\xi_t$,
and the logarithmic flux, $\log{\cal F}_x$, in the $uvbyUBVRIJHK$ passbands.

\begin{table*}
\caption{Tables of the passband fluxes, their derivatives over effective temperature and gravity, and
non-linear limb darkening coefficients, for the BSTAR2006 NLTE models in the $uvbyUBVRIJHK$ passbands.}             
\label{strumienie}      
\centering                          
\begin{tabular}{lllccccc}        
\hline\hline                 
 &  & & &  & & & \\
fluxes & derivatives & limb darkening & range of $T_{\rm eff}$ [K]& range of $\log g $ [dex]& $(Z/Z_{\odot})_{\rm atm}$ & $\xi_t$ [km/s] & mixture \\
 &  & coefficients & &  & &  & \\  
\hline                        
   flux\_BCv2 & der\_BCv2 & LDC\_BCv2 & 15000-30000 & 1.75-4.75  & 2.0 & 2 & GS98 \\
   flux\_BCv10 & der\_BCv10 & LDC\_BCv10 & 15000-30000 & 1.75-3.00  & 2.0 & 10 & GS98 \\
   flux\_BCv10CN & der\_BCv10CN & LDC\_BCv10CN & 15000-30000 & 1.75-3.00  & 2.0 & 10 & CN \\
   flux\_BGv2 & der\_BGv2 & LDC\_BGv2 & 15000-30000 & 1.75-4.75  & 1.0 & 2 & GS98 \\
   flux\_BGv10 & der\_BGv10 & LDC\_BGv10 & 15000-30000 & 1.75-3.00  & 1.0 & 10 & GS98 \\
   flux\_BGv10CN & der\_BGv10CN & LDC\_BGv10CN & 15000-30000 & 1.75-3.00  & 1.0 & 10 & CN \\
   flux\_BLv2 & der\_BLv2 & LDC\_BLv2 & 15000-30000 & 1.75-4.75  & 0.5 & 2 & GS98 \\
   flux\_BLv10 & der\_BLv10 & LDC\_BLv10 & 15000-30000 & 1.75-3.00  & 0.5 & 10 & GS98 \\
   flux\_BLv10CN & der\_BLv10CN & LDC\_BLv10CN & 15000-30000 & 1.75-3.00  & 0.5 & 10 & CN \\
   flux\_BSv2 & der\_BSv2 & LDC\_BSv2 & 15000-30000 & 1.75-4.75  & 0.2 & 2 & GS98 \\
   flux\_BSv10 & der\_BSv10 & LDC\_BSv10 & 15000-30000 & 1.75-3.00  & 0.2 & 10 & GS98 \\
   flux\_BSv10CN & der\_BSv10CN & LDC\_BSv10CN & 15000-30000 & 1.75-3.00  & 0.2 & 10 & CN \\
   flux\_BTv2 & der\_BTv2 & LDC\_BTv2 & 15000-30000 & 1.75-4.75  & 0.1 & 2 & GS98 \\
   flux\_BTv10 & der\_BTv10 & LDC\_BTv10 & 15000-30000 & 1.75-3.00  & 0.1 & 10 & GS98 \\
   flux\_BTv10CN & der\_BTv10CN & LDC\_BTv10CN & 15000-30000 & 1.75-3.00  & 0.1 & 10 & CN \\
   flux\_BZv2 & der\_BZv2 & LDC\_BZv2 & 15000-30000 & 1.75-4.75  & 0.0 & 2 & GS98 \\
   flux\_BZv10 & der\_BZv10 & LDC\_BZv10 & 15000-30000 & 1.75-3.00  & 0.0 & 10 & GS98 \\
\hline                                   
\end{tabular}
\tablefoot{
GS98 - chemical mixture by Grevesse \& Sauval (\cite{grevese}),
CN - chemical mixture enriched in helium and nitrogen, and depleted in carbon.}
\end{table*}

The most important atmospheric parameters in the expression for the brightness variation of a pulsating star
are the flux derivatives over effective temperature, $T_{\rm eff}$, and gravity, $\log g$.
In a given photometric pasband $x$, they are defined as
$$\alpha_T^x=\frac{\partial\log{\cal F}_x}
{\partial\log T_{\rm eff}}~~~{\rm and}~~~\alpha_g^x=\frac{\partial\log{\cal F}_x}
{\partial\log g}.$$
Values of these flux derivatives depend not only on effective temperature and gravity but also on the metallicity,
microturbulent velocity and the departure from LTE in the star's atmosphere.

Similarly to the passband fluxes, the derivatives were computed in the whole range of parameters
of the BSTAR2006 models. Names of the derivative tables are given in Table\,\ref{strumienie}.
Each file contains the following columns: line number, effective temperature, $T_{\rm eff}$,
logarithm of the surface gravity, $\log g$, metallicity, $(Z/Z_{\odot})_{\rm atm}$, microturbulent velocity, $\xi_t$,
and the flux derivatives over effective temperature, $\alpha_T$, and gravity, $\alpha_g$, in the $uvbyUBVRIJHK$ passbands.

In Fig.\,1, we show the NLTE flux derivatives as a function of temperature for two Str\"omgren passbands $uy$
and three values of gravity, $\log g=3.5, 4.0, 4.5$.
In the left panel, we plot the temperature derivative, $\alpha_T$,
and in the right one the gravity derivative, $\alpha_g$.  We assumed the solar metallicity, $(Z/Z_{\odot})_{\rm atm}$=1
and the microturbulent velocity of $\xi_t$=2 km/s.
As we can see the lower the gravity the larger values of $\alpha_T$ and the lower values of $\alpha_g$.

The wavelength dependence of the flux derivatives is presented in Fig.\,2.
We considered the central wavelengths, $\lambda_c$, of the $UBVRIJHK$ passbands and a model
with $T_{\rm eff}=20 000$ K and $\log g$ =4.0. The most steep derivatives are in the ultraviolet filter
and then the absolute values of $\alpha_T$ and $\alpha_g$ decrease quite rapidly with $\lambda_c$.
This is because the early B-type stars emit the most amount of their energy in the UV wavelength range.
In Fig.\,2, we compare also the NLTE derivatives with the LTE ones.
The largest difference can be seen again in the $U$ passband.

Effect of NLTE on the flux derivatives as a function of effective temperature for the Str\"omgren $uy$ passbands
is presented in Fig.\,3.  In general, differences between the LTE and NLTE derivatives are relatively small
and they increase with effective temperature, especially in the case of $\alpha_g$ in the $u$ passband.
In the case of both model atmospheres, the values of $\alpha_T$ in the $u$ passband change only slightly
with the temperature, whereas in the $y$ passband they are larger for higher $T_{\rm eff}$.
The gravity derivatives get more steep with the higher temperature, except $\alpha_g^u(T_{\rm eff})$
for NLTE models, which is a decreasing function up to $T_{\rm eff}\approx 24000$ K and then it increases.
\begin{figure*}
\centering
\includegraphics[angle=-90, width=\textwidth]{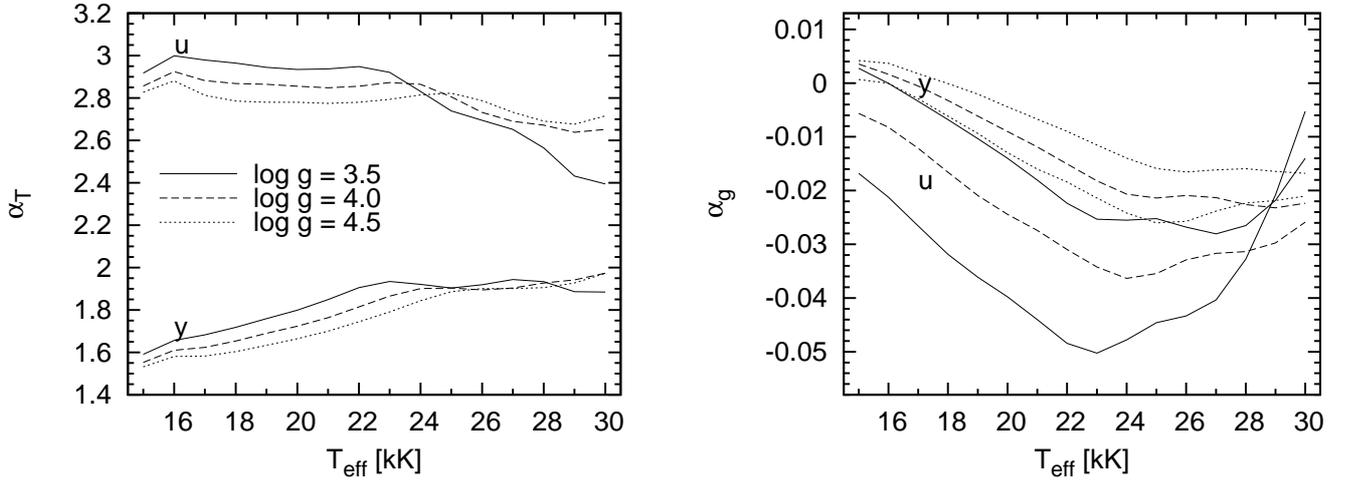}
\caption{The NLTE flux derivatives over $\log T_{\rm eff}$ (the left panel) and $\log g$ (the right panel) as a function of temperature
for the Str\"omgren $uy$ passbands. Three values of $\log g=3.5, 4.0, 4.5$ were considered. The values of the atmospheric metallicity
and microturbulent velocity were assumed as $(Z/Z_{\odot})_{\rm atm}$=1 and $\xi_t$=2 km/s, respectively.}
\label{der_3logg}
\end{figure*}

\begin{figure*}
\centering
\includegraphics[angle=-90, width=\textwidth]{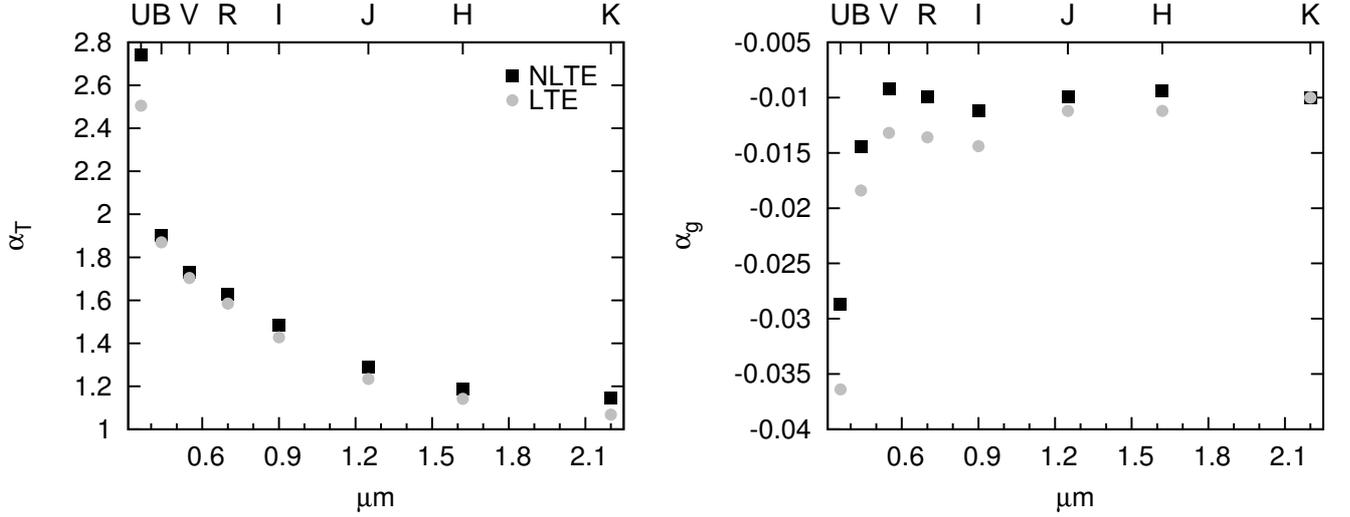}
\caption{A comparison of NLTE and LTE flux derivatives as a function of the central wavelengths of the $UBVRIJHK$ passbands,
for a models with $T_{\rm eff}$=20000 K, $\log g=4.0$, $(Z/Z_{\odot})_{\rm atm}$=1, $\xi_t$=2 km/s.
The left and right panel correspond to the temperature ($\alpha_T^x$) and gravity ($\alpha_g^x$) derivatives, respectively.}
\label{der_NLTE_LTEvUBVRIJHK}
\end{figure*}

\begin{figure*}
\centering
\includegraphics[angle=-90, width=\textwidth]{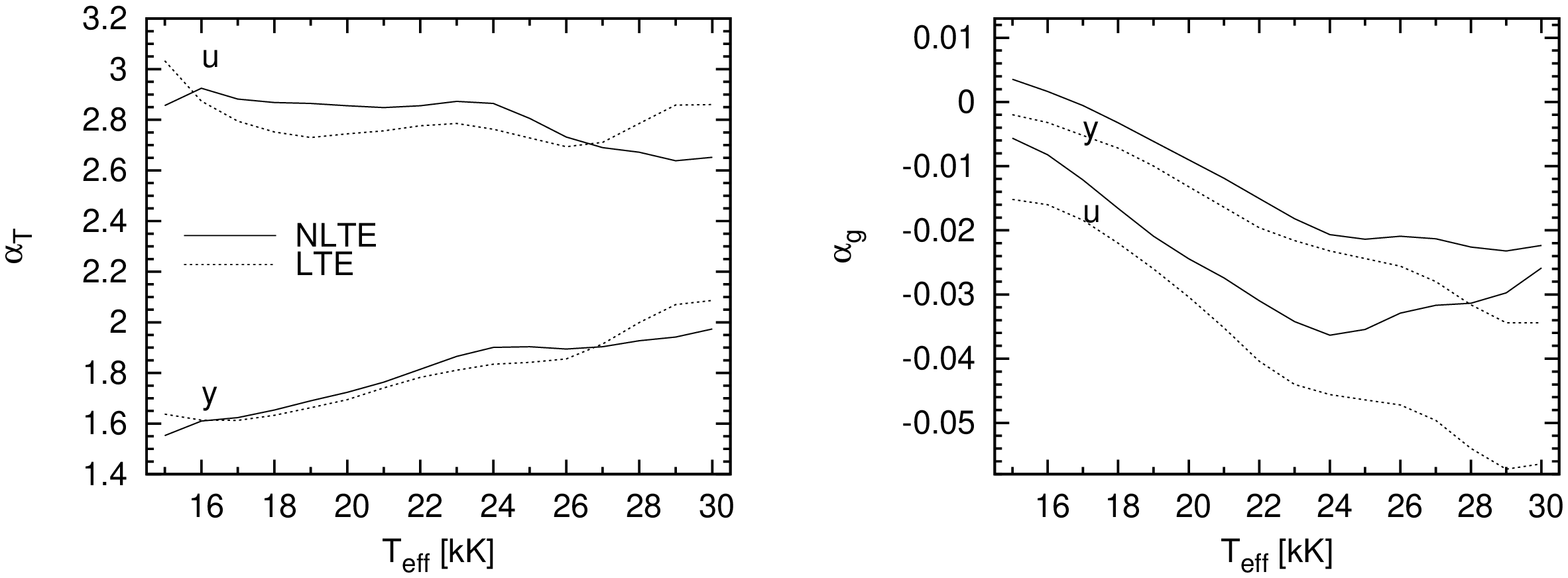}
\caption{Values of $\alpha_T^x$ (the left panel) and $\alpha_g^x$ (the right panel) as a function of temperature
 for LTE and NLTE models and the Str\"omgren $uy$ passbands.
The remaining parameters of the model are: $\log g=4.0$, $(Z/Z_{\odot})_{\rm atm}$=1, $\xi_t$=2 km/s.}
\label{der_NLTE_LTEvTeff}
\end{figure*}


\subsection{Limb darkening}
Knowledge of a distribution of the specific intensity over the stellar disc
is crucial in many fields of astrophysics.
In the case of pulsating stars, limb-darkening function and its derivatives
over effective temperature and gravity, are the second atmospheric data
needed to compute theoretical values of the photometric amplitudes and phases.

A usage of analytical formulae for the specific intensity distribution saves enormously the computation time.
The first limb-darkening law was proposed by Milne (\cite{milne}) in a linear form.
With the development of models of stellar atmospheres, it turned out that this approximation has a poor accuracy.
Thereafter, Klinglesmith \& Sobieski (\cite{kinglesmith}) published a logarithmic law for early type stars.
A quadratic formula for the limb-darkening law was proposed by Manduca at al. (\cite{manduca}) and Wade \& Ruci\'nski (1985).
For stars hotter than 8500 K, Di\'az-Cordov\'es \& Gim\'enez (\cite{diaz}) suggested a square root law.

However, all these above formulas were not adequate for all types of stars.
A more general law was proposed by Claret (\cite{claret2000}) in the following non-linear form
\begin{equation}
\label{nnclaret}
\frac{I(\lambda,\mu)}{I(\lambda,1)}=1-\sum_{k=1}^4a^\lambda_k(1-\mu^{\frac{k}{2}})
\end{equation}
where $I(\lambda,\mu)$ is a specific intensity at the wavelength $\lambda$,  $\mu$ is a cosine of the angle
between a line of sight and the emergent intensity, and $I(\lambda,1)$ is the value at the center of the stellar disc.
Limb darkening coefficients, $a^\lambda_k$, are determined to reproduce the model intensity distribution and to conserve
the flux with a high accuracy in the whole range of effective temperatures and gravities.

These coefficients were calculated for LTE model atmospheres in many photometric systems and
a wide range of effective temperatures, gravities, metallicities and for several values
of the microturbulent velocities (Claret 2000, 2003, Claret \& Hauschildt 2003, Claret 2008).
A comparison of different numerical methods of computations of limb darkening coefficients (LDC)
were discussed in detail by D\'iaz-Cordov\'es et al.\,(\cite{diaz1995}), Claret (\cite{claret2000}),
Heyrovsk\'y (\cite{heyrovsky}) and Claret (\cite{claret2008}).

In this paper, we determine non-linear LDC for metal line-blanketed, NLTE, plane-parallel, hydrostatic model atmospheres
of Lanz \& Hubeny (\cite{lanzhubeny}). In the first step, we computed monochromatic specific intensities
for 20 equally separated points of $\mu$ in the range of $\left< 0.001,1\right >$
and in the wavelength range of (2950,26500) $\AA$, for all parameters of the BSTAR2006 atmospheres.
All calculations were performed using the SYNSPEC program (Hubeny at al.\,\cite{synspec2}, Hubeny \& Lanz \cite{synspec}).
This code is intended to compute specific intensities and  fluxes from the model atmosphere input,
which we took from NLTE models described above. Subsequently, for each disc position, $\mu$,
we integrated intensities over the $uvby$ and $UBVRIJHK$ bands
\begin{equation}
\label{integ1}
I_x(\mu)=\frac{\int\limits^{\lambda_1}_{\lambda_2}I(\lambda,\mu)S(\lambda)\,d\lambda}{\int\limits^{\lambda_1}_{\lambda_2}S(\lambda)\,d\lambda},
\end{equation}
where $I_x(\mu)$ is the specific intensity in the passband $x$ and $I(\lambda, \mu)$ is the monochromatic
specific intensity obtained by SYNSPEC. The response function, $S(\lambda)$, was interpolated to the wavelengths
of computed monochromatic specific intensity by the cubic spline method.

The non-linear limb darkening coefficients, $a_k^x$, were determined using the least squares method by minimizing
\begin{equation}
\label{lastsquare}
\chi^2(a_1,a_2,a_3,a_4)=\sum_{i=1}^{20}\left(I_{LDC_i}-I_i\right)^2,
\end{equation}
where $I_{LDC_i}$ is the specific intensity computed with the Claret nonlinear law
and $I_i$ is the corresponding model intensity, at the point $\mu_i$.

Similarly to Heyrovsk\'y (\cite{heyrovsky}), we evaluated the quality of our fits by the relative residuals
\begin{equation}
\label{sigma}
\sigma^2=\frac{\sum\limits_{i=1}^{20}\left(I_i-I_{LDC_i}\right)^2}{\sum\limits_{i=1}^N I_i^2},
\end{equation}
whereas, the conservation of the flux was controlled by computing the relative flux excess
\begin{equation}
\label{flux_cons}
\frac{\Delta {\cal F}_x} {{\cal F}_x} =\frac{\int \limits_0^1 I_x \mu \,d\mu-\int \limits_0^1 I_{x,LDC} \mu \,d\mu}{\int \limits_0^1 I_x \mu \,d\mu}.
\end{equation}

The limb darkening law, $h(\mu)$, given in Eq.\,\ref{eq5}, is defined as (Daszy\'nska-Daszkiewicz et al. 2002)
\begin{equation}
h_x(\mu)=2\pi~\frac{I_x(\mu)}{{\cal F}_x}=2\pi~\frac{1 -\sum\limits_{k=1}^4 a_k^x (1
-\mu^{k/2})} {1-\sum\limits_{k=1}^4 \frac{k}{k+4} a_k^x }.
\end{equation}
%

Names of the LDC files are given Table\,\ref{strumienie} and they are coded in the same way as the flux
and flux derivatives tables (cf. Sect.\,3.1). Each file contains the following columns: names of LDC, $a_k$,
effective temperature, $T_{\rm eff}$, gravity, $\log g$, metallicity, $(Z/Z_{\odot})_{\rm atm}$, mictroturbulent velocity, $\xi_t$,
and values of LDC in the Str\"{o}mgren and Johnson-Cousins-Glass photometric passbands, in the order: $uvbyUBVRIJHK$.

In Fig.\,4 we show distributions of the normalized specific intensity in the $UBVRIJHK$ passbands for
a model with the following parameters: $T_{\rm eff}$=20000 K, log g=4.0, $\xi _t$=2 km/s, $(Z/Z_{\odot})_{\rm atm}$=1.
In this figure, we compare the specific intensity computed by means of SYNSPEC for 20 equally separated points
of $\mu$ (squares) and the fitted limb darkening function defined by Eq.\,\ref{nnclaret} (solid lines).
\begin{figure}
\centering
\includegraphics[angle=-90, width=\columnwidth]{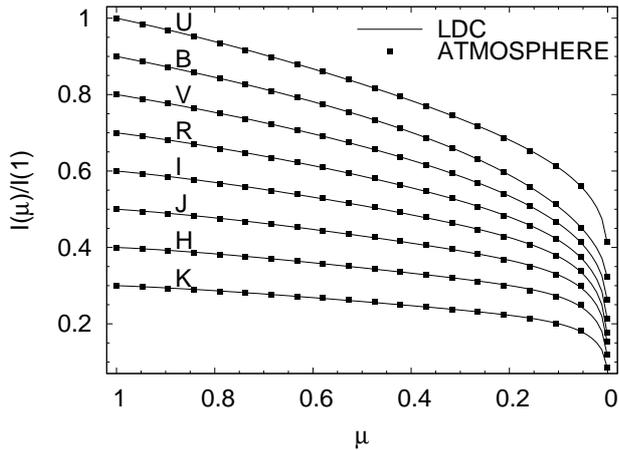}
\caption{The angular distribution of the normalized specific intensity for the
         $UBVRIJHK$ passbands  and the NLTE model with $T_{\rm eff}$=20000 K,
         $\log g$=4.0, $\xi _t$=2 km/s and $(Z/Z_{\odot})_{\rm atm}$=1.
         The actual model intensities are marked as squares and intensities computed
         from LDC as solid lines. For clarity, passbands from $B$ are shifted downward by 0.1.}
\label{LDCatmospheres}
\end{figure}
Intensities for passbands $BVRIJHK$ were shifted downward by $n\cdot 0.1$, where n=1,2,...,7.
As we can see a quality of the limb darkening fit is very accurate.

To compare values of the NLTE and LTE intensities, in Fig.\,5 we plot $I_x$ [erg$\cdot$cm$^{-2}\cdot$s$^{-1}\cdot$ster$^{-1}$]
as a function of the angle $\mu$ for the Str\"omgren $uvby$ passbands. The same model as in Fig.\,4 was considered.
The NLTE intensities get lower values, in particular, for the $u$ passband.
This is caused by the location of the $u$ filter on the Balmer continuum where
the difference in the amount of energy radiated in LTE and NLTE models is more pronounced
than in a region to the right from the Balmer jump, where other filters ($vby$) are defined.
\begin{figure}
\centering
\includegraphics[angle=-90, width=\columnwidth]{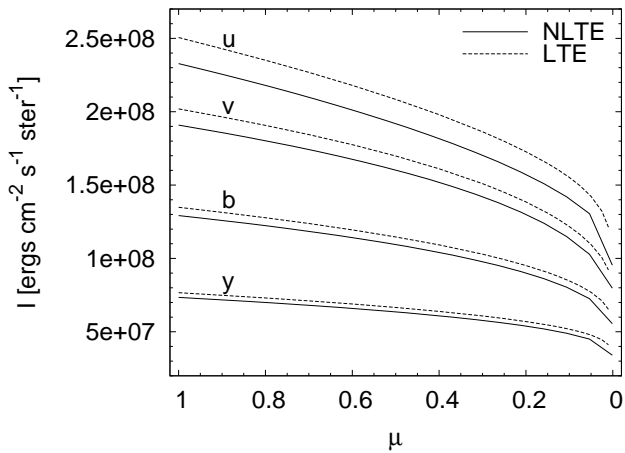}
\caption{The intensity distribution for the $uvby$ passbands for
         the NLTE (solid lines) and LTE (dashed lines) models
         with  $T_{\rm eff}$=20000 K, $\log g$=4.0, $\xi_t$=2 km/s and $(Z/Z_{\odot})_{\rm atm}$=1.}
\label{int_T-K}
\end{figure}

In Fig.\,6 and 7, we plot the NLTE limb darkening coefficients, $a_1, a_2, a_3, a_4$,
as a function of $T_{\rm eff}$ in the Str\"omgren $u$ passband, for different values of gravity, $\log g$,
and metallicity, $(Z/Z_{\odot})_{\rm atm}$, respectively. All panels have the same scale.
As we can see, there is a strong dependence of LDC
on effective temperature, gravity and metallicity. The sensitivity on $T_{\rm eff}$ is stronger
for lower values of $\log g=3.5$. The sensitivity to metallicity is similar for all LDC.
The effect of the microturbulent velocity, $\xi_t$,
not shown here, is comparable to the effect of metallicity.
\begin{figure*}
\centering
\includegraphics[angle=-90, width=\textwidth]{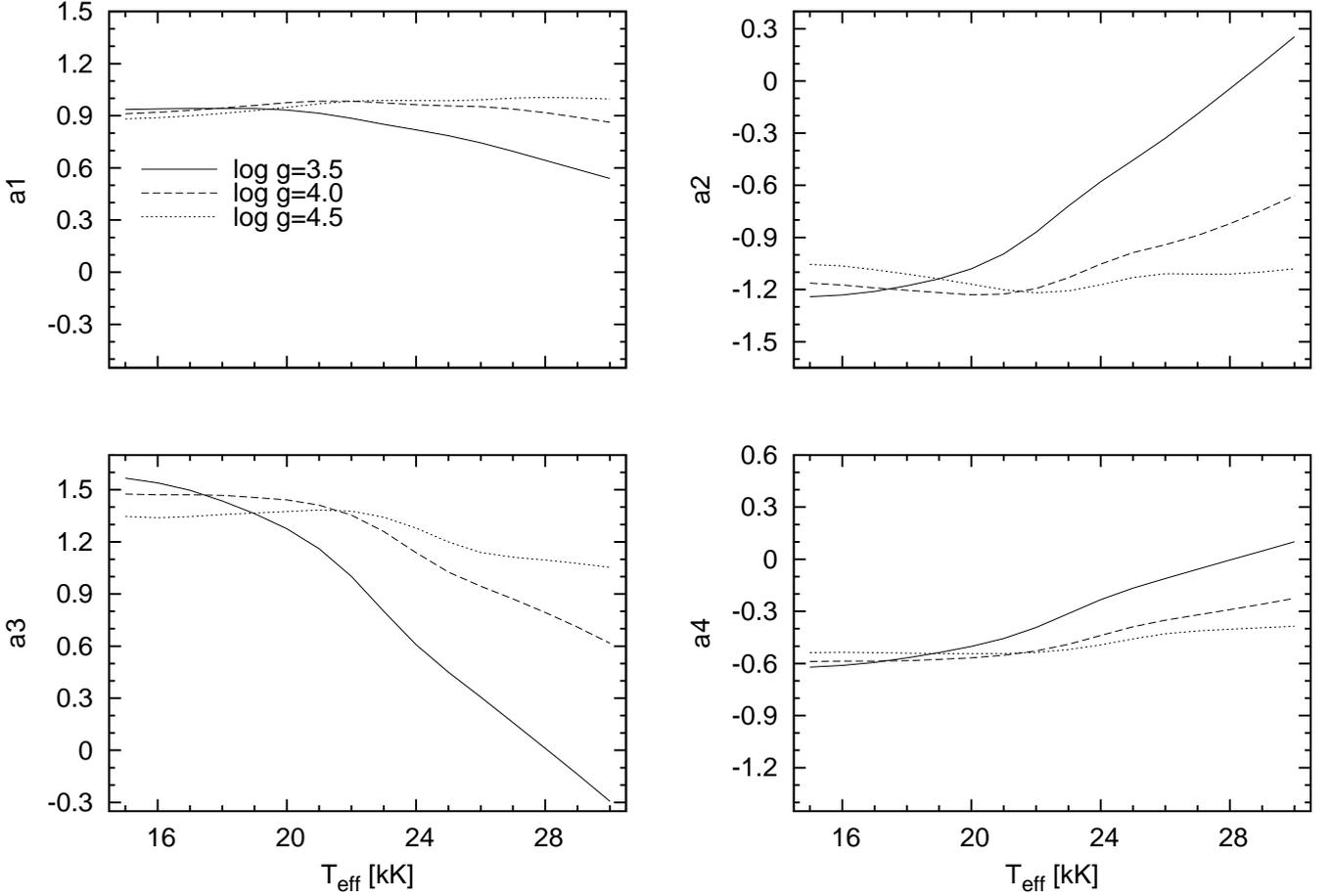}
\caption{The NLTE limb darkening coefficients in the $u$ passband as a function of $T_{\rm eff}$
for three values of gravity: $\log=3.5$ (solid line), $\log=4.0$ (dashed line) and $\log=4.5$ (dotted line).
The solar metallicity, $(Z/Z_{\odot})_{\rm atm}$=1, and microturbulent velocity  of $\xi_t$=2 km/s were assumed.}
\label{LDClogg}
\end{figure*}
\begin{figure*}
\centering
\includegraphics[angle=-90, width=\textwidth]{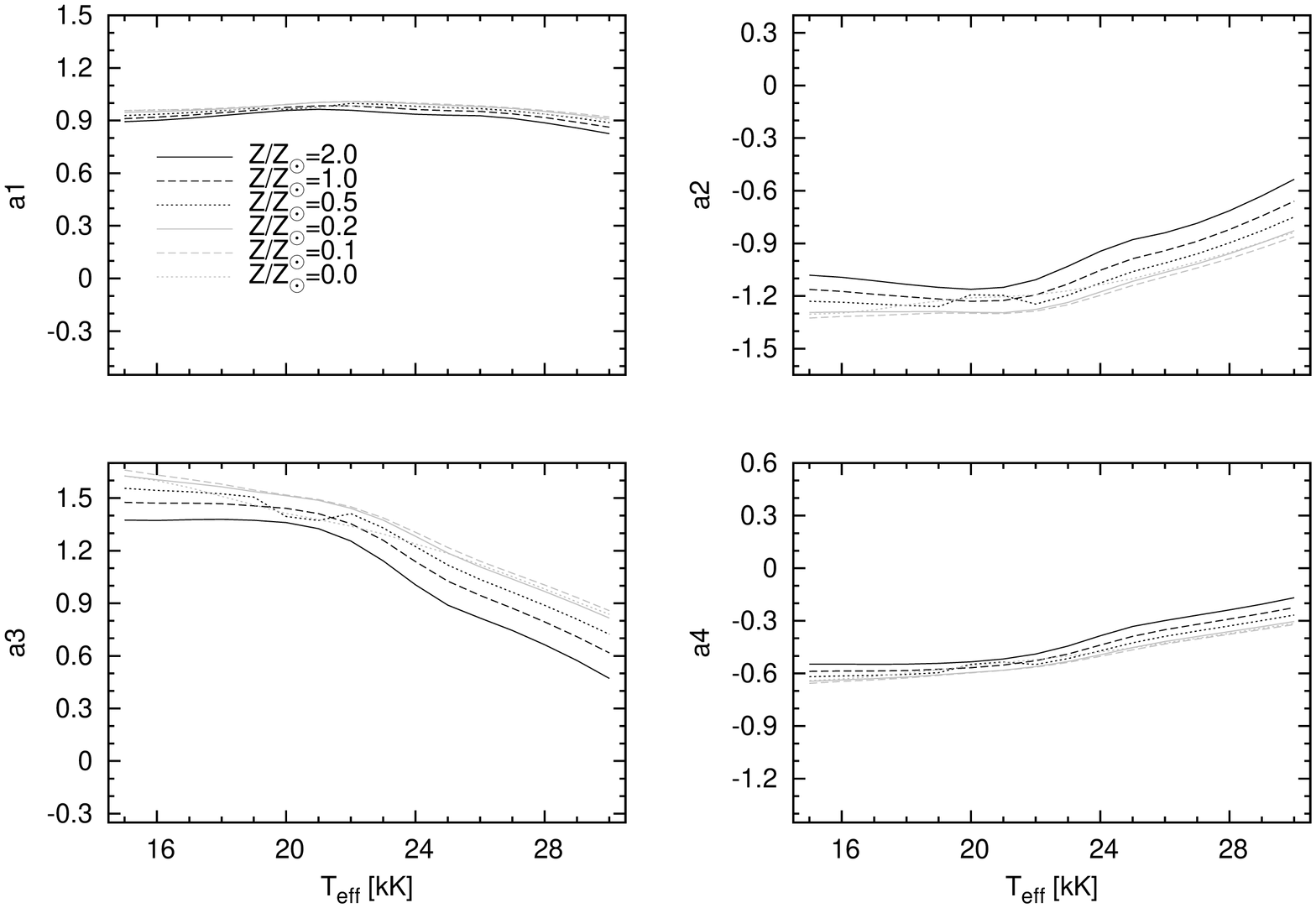}
\caption{The NLTE limb darkening coefficients in the $u$ passband  as a function of $T_{\rm eff}$
for six values of the atmospheric metallicity: $(Z/Z_{\odot})_{\rm atm}$=2.0 (solid black line), 1.0 (dashed black line),
          0.5 (dotted black line), 0.2 (solid grey line), 0.1 (dashed grey line) and 0.0 (dotted grey line).
          The gravity of $\log g$=4.0 and microturbulent velocity of $\xi_t$=2 km/s were assumed.}
\label{LDCzalmet}
\end{figure*}

Let us now discuss the accuracy of our determinations of the Claret limb-darkening coefficients
for the BSTAR2006 NLTE models.
In Table.\,2, we give the average and maximum values of the merit functions,
i.e. the relative residuals, $\sigma$, and relative flux excess, $\left|\Delta F \over F \right|$.
These values were calculated for the whole range of $T_{\rm eff}$, i.e. (15000,30000) K, and $\log g$, i.e. (1.75, 4.75) dex,
assuming the solar metallicity, $(Z/Z_{\odot})_{\rm atm}$=1, and the microturbulent velocity of $\xi_t$=2 km/s.
We used 20 equally spaced points of $\mu$ instead of 17 unequally spaced points as used in LTE Kurucz models.
Moreover, our NLTE values of $I(\mu)$ were computed closer to the stellar limb, i.e. up to $\mu=0.001$,
whereas the lowest angle of $\mu$ in LTE models is 0.01.
We considered two sets of passbands: $uvbyUBVRIJHK$ and $BVRI$.
The second set was used to compare our results with those of Heyrovsk\'y (\cite{heyrovsky})
who made computations for LTE Kurucz atmosphere models using 17 and 11 points of $\mu$.
As we can see the quality of our fit is very good. The average value of $\sigma$ amounts
to 0.119\% and for the worst fitted profile we got $\sigma$=0.432\%.
Our average value of $\sigma$ is smaller than the Heyrovsk\'y's one by a factor of two.
When we limited our analysis to the $BVRI$ passbans, as in Heyrovsk\'y (\cite{heyrovsky}),
the result is even better. In this case, the average and maximum values of $\sigma$
amount to 0.089\% and 0.251\%, respectively.
This indicates that with the larger number of equally separated points of $\mu$
in the fitting procedure one reproduces more accurately the model intensities.

\begin{table}
\caption{Comparison of our limb-darkening fit quality with results of Heyrovsky (2007)
         for the Claret non-linear law. In columns 2 and 3, we give the average and maximum values
         of the relative rms residual, $\sigma$, respectively. Columns 4 and 5 contain
         the average and maximum values of the relative flux excess, $\left|\Delta F \over F \right|$.
         Passbands for which these goodness-of-fit quantities were evaluated are given in notes.}             
\label{por_sig}      
\centering                          
\begin{tabular}{ccccc}        
\hline\hline                 
Method & Average $\sigma$ & Max. $\sigma$ & Average $\left|\Delta F \over F \right| $ & Max. $\left|\Delta F \over F \right| $ \\
&   \% & \% \\    
\hline                        
   20-point$^a$ & 0.119  & 0.432 & $14.5\times 10^{-5}$  & $9.10\times 10^{-4}$ \\
   20-point$^b$ & 0.089 & 0.251 & $9.50\times 10^{-5}$ & $5.28\times 10^{-4}$ \\
   17-point$^c$ & 0.190 & 0.625 & $8.27\times 10^{-5}$ & $5.05\times 10^{-4}$ \\      
   11-point$^c$ & 0.175 & 0.558 & $4.57\times 10^{-5}$ & $2.17\times 10^{-4}$ \\
\hline                                   
\end{tabular}
\tablefoot{
$^a$ values evaluated for all 12 passbands: $uvbyUBVRIJHK$.
$^b$ values evaluated for the $BVRI$ passbands.
$^c$ values from Heyrovsk\'y (\cite{heyrovsky}) for the $BVRI$ passbands.}

\end{table}

Finally, let us check the conservation of the flux.
The average value of $\left|\Delta F \over F \right| $ from our fitting
for NLTE models is slightly worse than in Heyrovsk\'y (\cite{heyrovsky})
for LTE models atmospheres, but of the same order of magnitude.
The maximum values of $\left|\Delta F \over F \right| $ are $9.10 \times 10^{-4}$ for all 12 passbnds
and $5.28 \times 10^{-4}$ for $BVRI$. The corresponding average values are $14.5 \times 10^{-5}$
and $9.50 \times 10^{-5}$, respectively.
For larger values of gravities, the fluxes computed with the Claret limb darkening law are overestimated
whereas for smaller values of $\log g$, fluxes are underestimated.
This is caused by a steeper slope of the intensity near the limb for lower values of $\log g$.
Consequently, the fitted limb darkening for lower gravities is slightly below model intensities.

\section{Uncertainties in photometric pulsational observables from model atmospheres}

\begin{figure*}
\centering
\includegraphics[angle=-90, width=\textwidth]{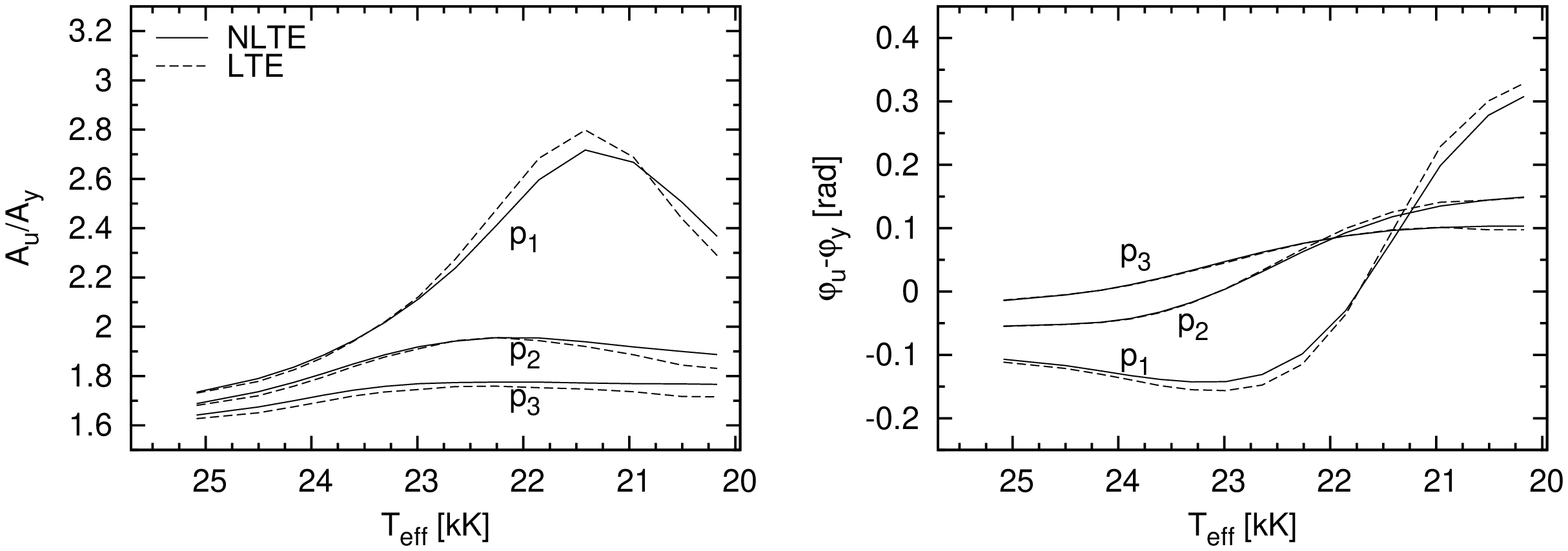}
\caption{A comparison of the NLTE and LTE values of the amplitude ratios, $A_u/A_y$ (the left panel)
         and phase differences,  $\varphi _u-\varphi_y$ (the right panel) for 10 $M_{\odot}$ main sequence models
         as a function of $T_{\rm eff}$ for the first three radial modes: p$_1$, p$_2$, p$_3$.
         The atmospheric  metallicity of $(Z/Z_{\odot})_{\rm atm}=1$ and microturbulent velocity of $\xi _t=2$ km/s were assumed.
         Linear nonadiabatic pulsations were computed with hydrogen and metal abundance of $X=0.7$ and $Z=0.02$, respectively,
         the OPAL opacities and the AGSS09 chemical mixture.
        }
\label{ef_atm_NLTE_LTEl0vTeff}
\end{figure*}
\begin{figure*}
\centering
\includegraphics[angle=-90, width=\textwidth]{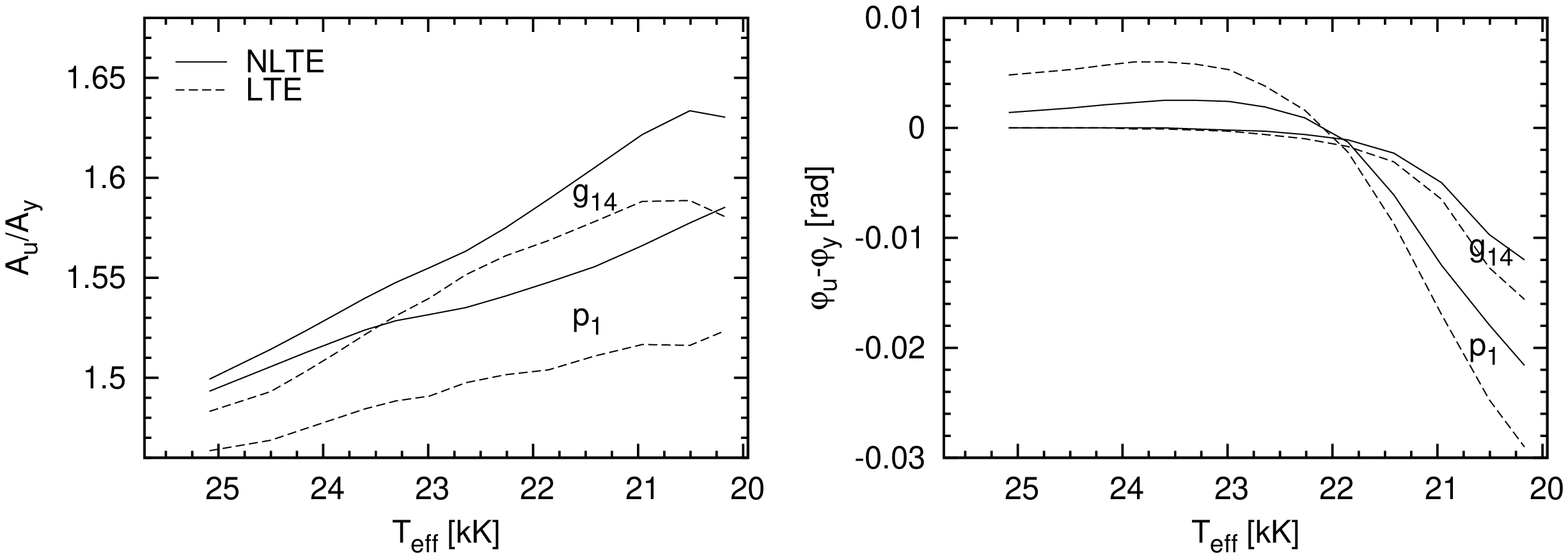}
\caption{The same as in Fig.\,\ref{ef_atm_NLTE_LTEl0vTeff} but for two $\ell=1$ modes: p$_1$ and g$_{14}$.}
\label{ef_atm_NLTE_LTEl1vTeff}
\end{figure*}
\begin{figure*}
\centering
\includegraphics[angle=-90, width=\textwidth]{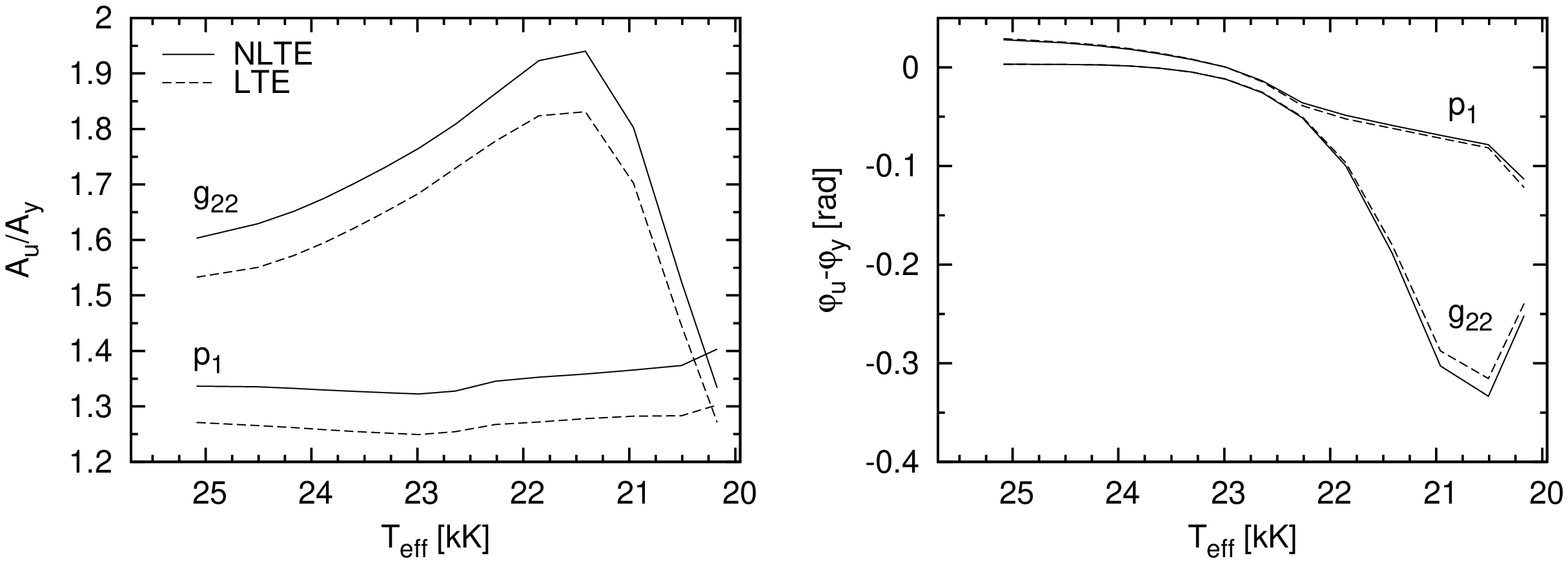}
\caption{The same as in Fig.\,\ref{ef_atm_NLTE_LTEl0vTeff} for for two $\ell=2$ modes: p$_1$ and g$_{22}$.}
\label{ef_atm_NLTE_LTEl2vTeff}
\end{figure*}

\begin{figure*}
\centering
\includegraphics[angle=-90, width=\textwidth]{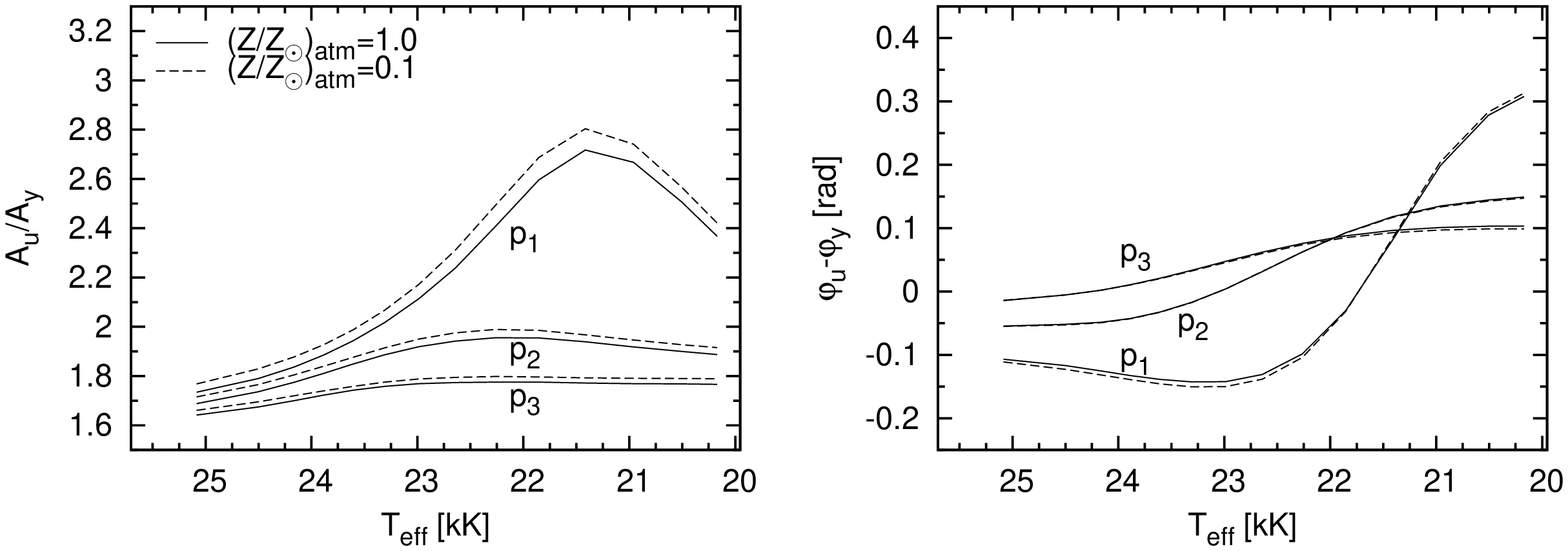}
\caption{Effect of the atmospheric metallicity on the amplitude ratios, $A_u/A_y$ (the left panel)
         and phase differences,  $\varphi _u-\varphi_y$ (the right panel) for 10 $M_{\odot}$ main sequence models
         as a function of $T_{\rm eff}$ for three radial modes: p$_1$, p$_2$, p$_3$.
         NLTE model atmospheres with the microturbulent velocity of $\xi _t=2$ km/s and two values of
         $(Z/Z_{\odot})_{\rm atm}$ were used.
        }
\label{ef_atm_NLTE_metl0vTeff}
\end{figure*}

The atmospheric metallicity, $(Z/Z_{\odot})_{\rm atm}$, microturbulent velocity, $\xi_t$, and effects of NLTE
are the most important factors which can affect the photometric amplitudes and phases of a pulsating star.

In all comparisons, we used a reference model computed with
the mass of $M=10M_{\odot}$, the OPAL opacity tables, the AGSS09 mixture, hydrogen abundance of $X$=0.7,
metal abundance of $Z$=0.02, without overshooting from a convective core, $\alpha_{\rm ov}$=0.0, and
NLTE-TLUSTY models of stellar atmospheres with the metallicity of $(Z/Z_{\odot})_{\rm atm}$=1.0 and
the microturbulent velocity of $\xi_t$=2 km/s.

In Fig.\,\ref{ef_atm_NLTE_LTEl0vTeff}, we compare the photometric observables computed with LTE and NLTE model atmospheres
at the same values of metallicity, $(Z/Z_{\odot})_{\rm atm}=1$, and microturbulent velocity, $\xi_t=2$ km/s,
for the $10 M_{\odot}$ model in the course of its main sequence evolution.
We considered the Str\"omgren $uy$ passbands and the first three radial modes: p$_1$, p$_2$, p$_3$.
In the left panel, we show the amplitude ratios, $A_u/A_y$,
and in the right one the corresponding phase differences, $\varphi_u-\varphi_y$, as a function of $T_{\rm eff}$.
As we can see, the fundamental mode is most sensitive to the NLTE effects.
In general, the NLTE values of the amplitude ratio,  $A_u/A_y$, are smaller than the LTE ones
for the hotter models and larger for more evolved models.

Subsequently we checked the NLTE effect for nonradial modes.
In Fig.\,\ref{ef_atm_NLTE_LTEl1vTeff}, we plot the same photometric observables as in Fig.\,\ref{ef_atm_NLTE_LTEl0vTeff},
but for the $\ell=1$ mode considering one pressure mode, p$_1$, and one high-order gravity mode, g$_{14}$.
Fig.\,\ref{ef_atm_NLTE_LTEl2vTeff} illustrates the same but for two $\ell=2$ modes: p$_1$ and g$_{22}$.
In the case of nonradial modes, the amplitude ratios, $A_u/A_y$, have larger values for NLTE computations.
We can see also a different behavior and values of $A_u/A_y$ and $\varphi_u-\varphi_y$
for pressure and high-order gravity modes, especially for the $\ell=2$ modes.

Effect of the atmospheric metallicity on the photometric observables for the radial modes are shown in Fig.\,\ref{ef_atm_NLTE_metl0vTeff}.
Here, we considered $(Z/Z_{\odot})_{\rm atm}=0.1$ in addition to our standard value of $(Z/Z_{\odot})_{\rm atm}=1.0$.
As we can see, the effect of the atmospheric metallicity is comparable to the effect of the departure from LTE in stellar atmospheres.
In the case of nonradial modes the effect of $(Z/Z_{\odot})_{\rm atm}$ is negligible.

Influence of the microturbulent velocity, $\xi_t$, is of the same order as $(Z/Z_{\odot})_{\rm atm}$,
as has been shown by Szewczuk \& Daszy\'nska-Daszkiewicz (2010).

\section{Uncertainties in pulsational photometric observables from pulsation theory}

Linear nonadiabatic theory of stellar pulsation provides eigenfrequencies, corresponding eigenfunctions
and information on mode excitation.
The value of the flux eigenfunction at the level of the photosphere enters the expression for the light variation
and it is called the $f$-parameter, as it has been introduced in Sect.\,2.
For a given mode frequency, this parameter depends, in general, on mean stellar parameters,
chemical composition, opacity data, subphotospheric convection etc.
Moreover, in the case of high-order gravity modes, the $f$-parameter depends also on the mode degree, $\ell$ (e.g. Daszy\'nska-Daszkiewicz \& Walczak 2010).
There are also large differences in values of $f$ between high and low frequency modes of a given degree, $\ell$.
This is why the amplitude ratios and phase differences for p- and high-order g-modes behave so different,
which is exemplified most pronounced in Fig.\,10.
\begin{figure*}
\centering
\includegraphics[angle=-90, width=\textwidth]{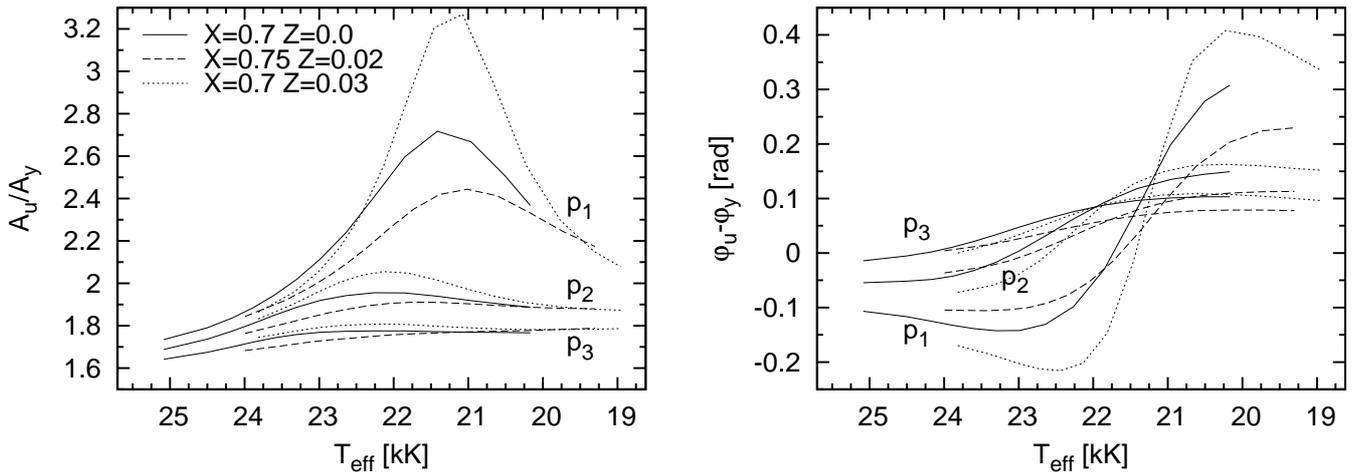}
\caption{Effects of the internal abundances of hydrogen, $X$, and metals, $Z$, on photometric observables for the first three radial modes as a function
         of $T_{\rm eff}$ for the 10 $M_{\odot}$ main sequence models.
         The NLTE models with the atmospheric metallicity of $(Z/Z_{\odot})_{\rm atm}=1$ and microturbulent velocity of $\xi _t=2$ km/s were used.
         Linear nonadiabatic pulsations were computed with OPAL opacity tables and AGSS09 chemical mixture.
        }
\label{ef_XZ_NLTE_l0vTeff}
\end{figure*}

\begin{figure*}
\centering
\includegraphics[angle=-90, width=\textwidth]{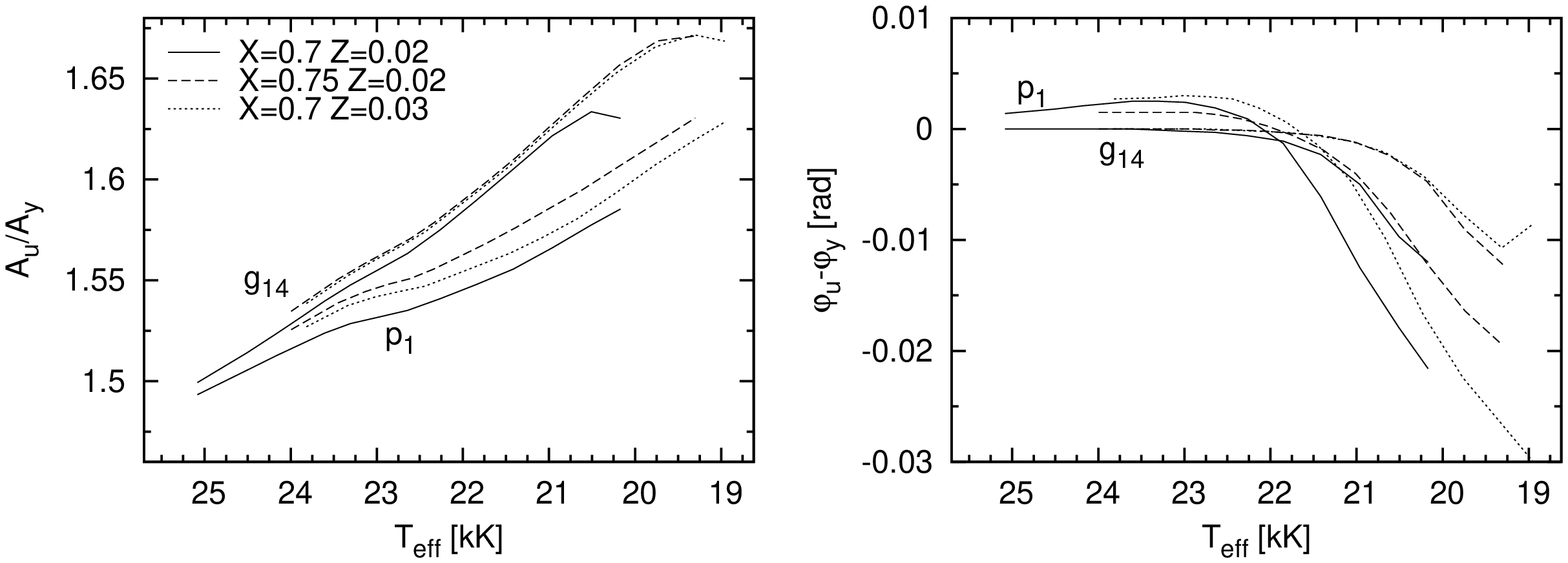}
\caption{The same as in Fig.\,\ref{ef_XZ_NLTE_l0vTeff} but for two $\ell=1$ modes: p$_1$ and g$_{14}$.}
\label{ef_XZ_NLTE_l1vTeff}
\end{figure*}

\begin{figure*}
\centering
\includegraphics[angle=-90, width=\textwidth]{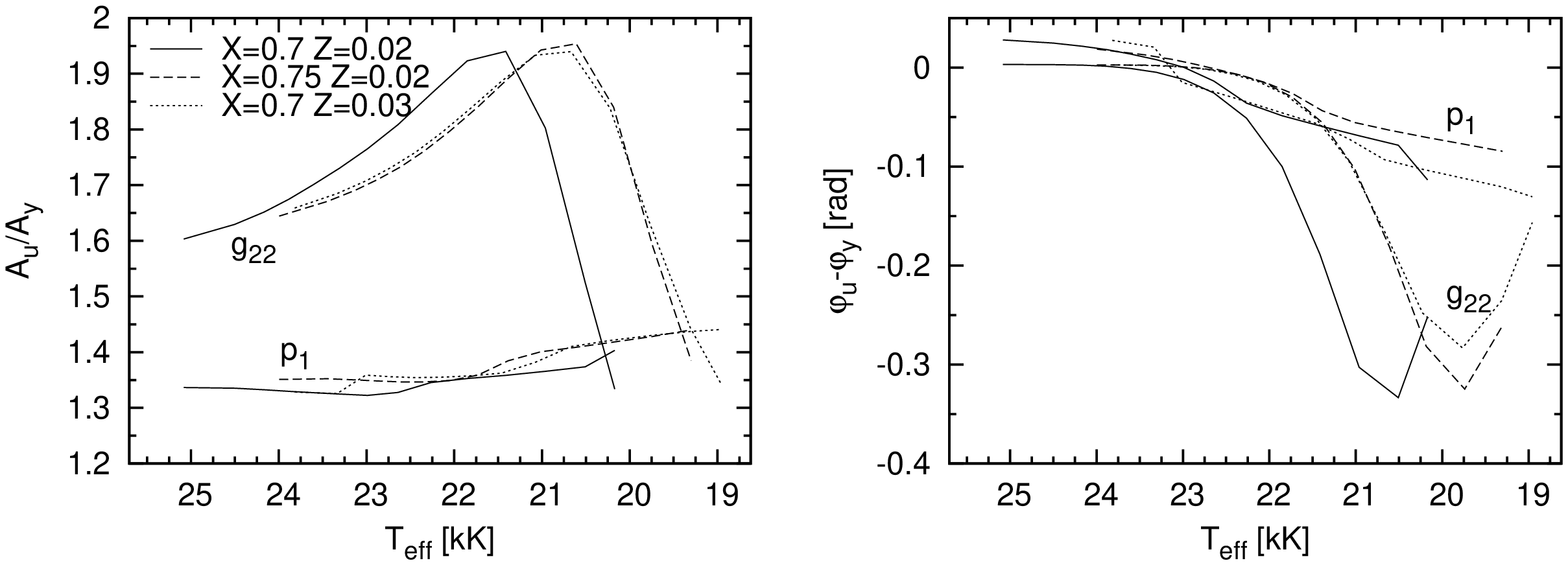}
\caption{The same as in Fig.\,\ref{ef_XZ_NLTE_l0vTeff} but for two $\ell=2$ modes: p$_1$ and g$_{22}$.}
\label{ef_XZ_NLTE_l2vTeff}
\end{figure*}

\begin{figure*}
\centering
\includegraphics[angle=-90, width=\textwidth]{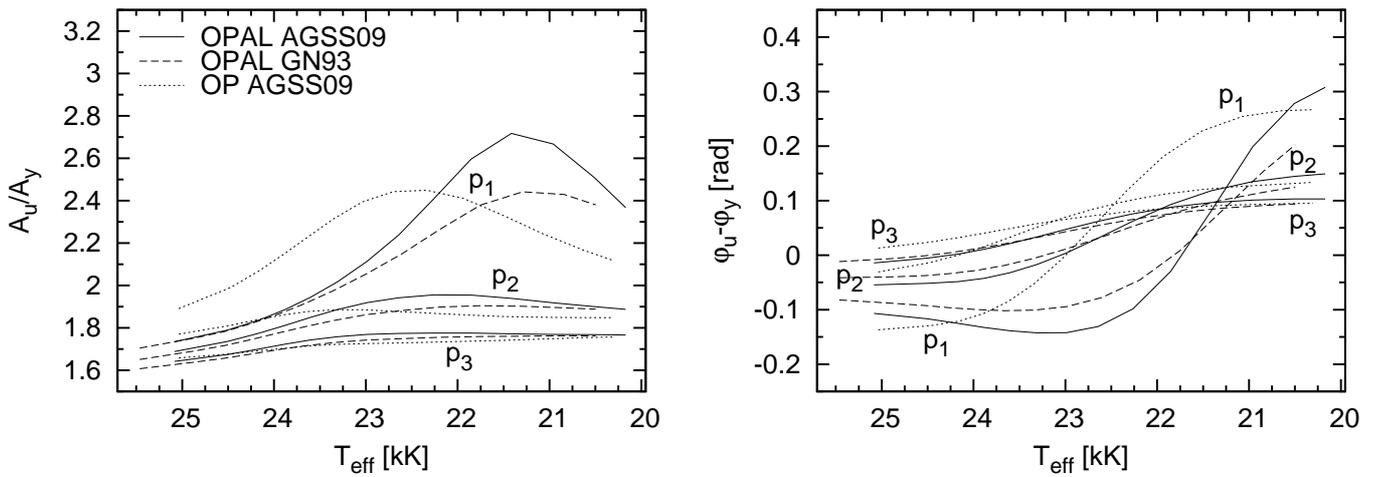}
\caption{Effects of the opacity tables and chemical mixture on photometric observables for the first three radial modes as a function
         of $T_{\rm eff}$ for the 10 $M_{\odot}$ main sequence models.
         The NLTE models with the metallicity of $(Z/Z_{\odot})_{\rm atm}=1$ and microturbulent velocity of $\xi _t=2$ km/s were used.
         Linear nonadiabatic pulsations were computed with $X=0.7$ and $Z=0.02$.}
\label{ef_tab_mix_NLTE_l0vTeff}
\end{figure*}

\begin{figure*}
\centering
\includegraphics[angle=-90, width=\textwidth]{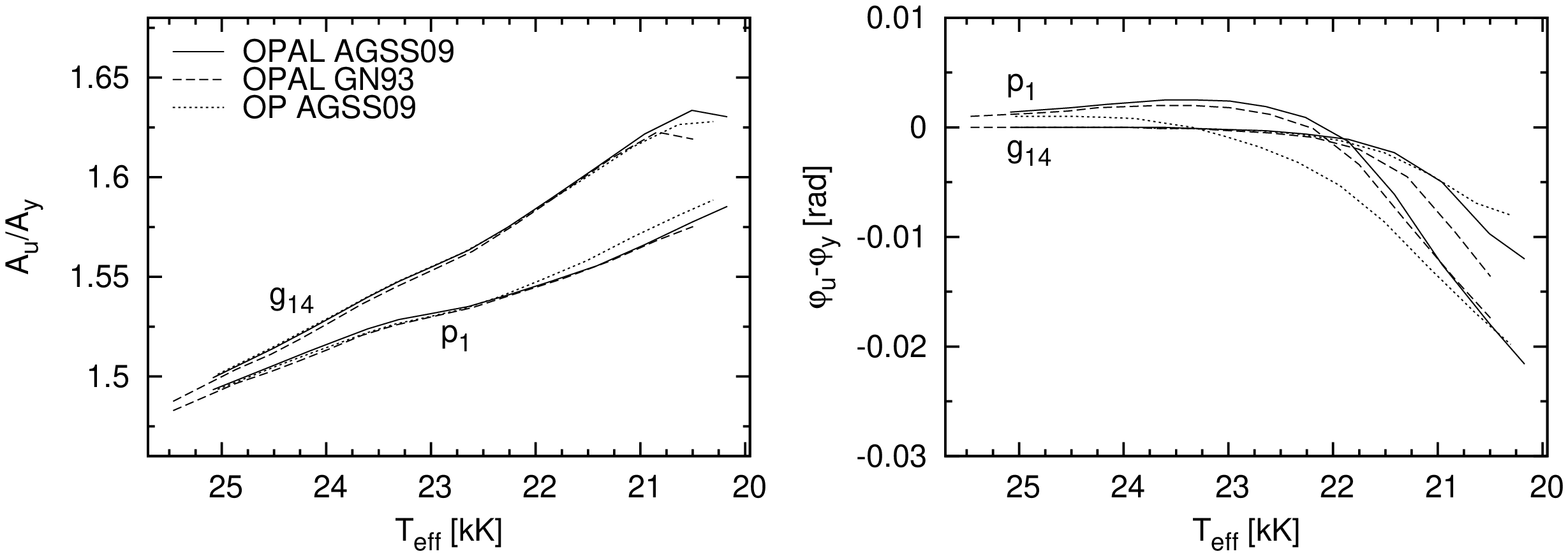}
\caption{The same as in Fig.\,\ref{ef_tab_mix_NLTE_l0vTeff} but for two $\ell=1$ modes: p$_1$ and g$_{14}$.}
\label{ef_tab_mix_NLTE_l1vTeff}
\end{figure*}

\begin{figure*}
\centering
\includegraphics[angle=-90, width=\textwidth]{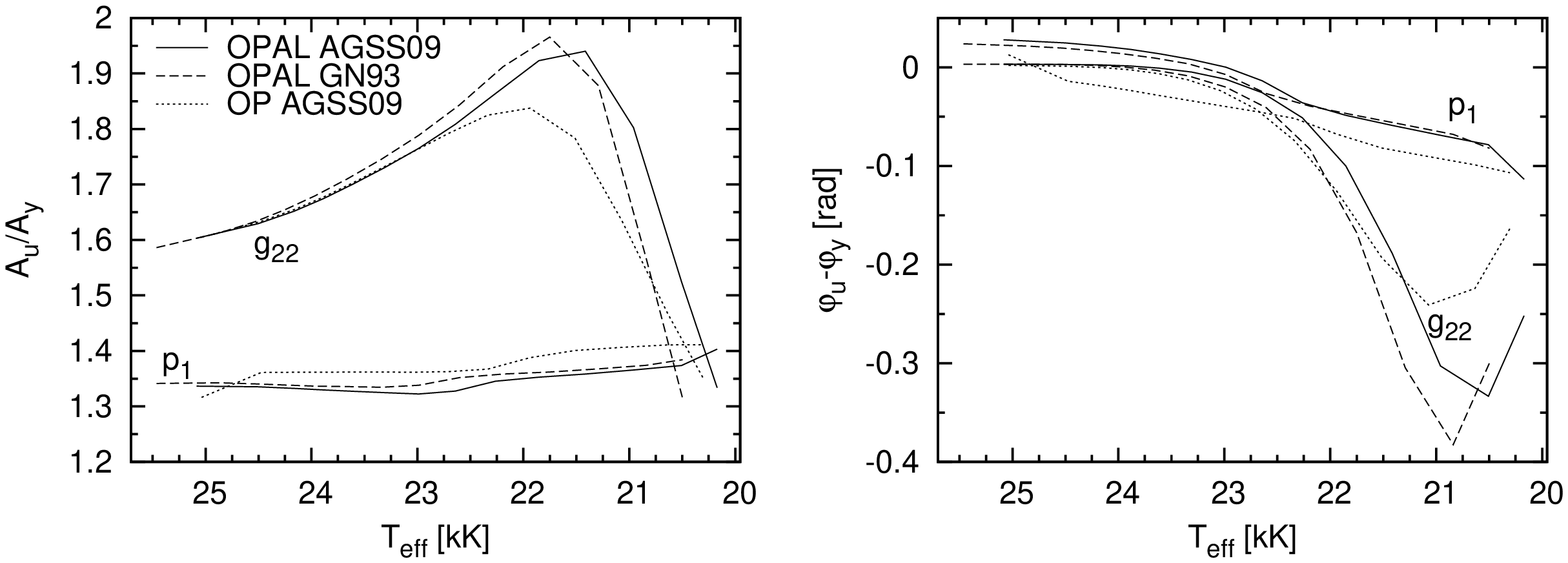}
\caption{The same as in Fig.\,\ref{ef_tab_mix_NLTE_l0vTeff} but for two $\ell=2$ modes: p$_1$ and g$_{22}$.}
\label{ef_tab_mix_NLTE_l2vTeff}
\end{figure*}

Pulsations of the B-type stars are driven by the $\kappa$ mechanism operating in the ,,Z-bump'' layer.
The nonadiabatic complex parameter, $f$, is also defined in this subphotospheric layer,
therefore its value strongly depends on metal abundances, chemical mixture and hence opacities
(Daszy\'nska-Daszkiewicz, Dziembowski, Pamyatnykh 2005).

In Fig.\,\ref{ef_XZ_NLTE_l0vTeff}, we show effects of the internal abundances of hydrogen, $X$, and metals, $Z$,
on the photometric amplitude ratios, $A_u/A_y$, and phase differences, $\varphi_u-\varphi_y$,
for the first three radial modes: p$_1$, p$_2$, p$_3$. The same model as in Sect.\,4 was selected.
Here, we compare computations obtained with $Z$=0.02 $vs.$ $Z$=0.03 and $X$=0.7 $vs.$ $X$=0.75.
As we can see, the amplitude ratios computed with $Z$=0.03 are larger than
those obtained with $Z$=0.02, whereas increasing hydrogen abundance, $X$, decreases
the amplitude ratios. The value of $\varphi_u-\varphi_y$ changes by about 0.1 rad at most.
These effects are particulary distinct for the radial fundamental mode, p$_1$.
Comparing Fig.\,\ref{ef_XZ_NLTE_l0vTeff} with Fig.\,8 and 11, we can see also that the effects of $X$ and $Z$
on the photometric observables are much more pronounced than the effects of NLTE and $(Z/Z_{\odot})_{\rm atm}$.
Please note that for a given mode degree, $\ell$, scales of the amplitude ratios
and phases differences are the same in all figures.

Effects of the hydrogen and metallicity abundance for the $\ell=1$, p$_1$ and $g_{14}$ modes is shown
in  Fig.\,\ref{ef_XZ_NLTE_l1vTeff}, and for the $\ell=2$, p$_1$ and g$_{22}$ modes in Fig.\,\ref{ef_XZ_NLTE_l2vTeff}.
As one can see, in the case of the nonradial modes these parameters have smaller influence on the photometric
observables than in the case of radial modes.
The amplitude ratio, $A_u/A_y$, for the $\ell=1$ mode increases with increasing abundances of both $Z$ and $X$.
This is true also for the $\ell=2$, p$_1$ mode, whereas the amplitude ratio of the $\ell=2$, g$_{22}$ mode
does not behave monotonically.
The important result is that in the case of nonradial modes the effects of $X$ and $Z$
are considerable smaller than the NLTE effects (cf. Fig.\,9 and 10), which is opposite to
the radial mode case.

Finally, we evaluated effects of the opacity data and chemical mixture.
In Fig.\,\ref{ef_tab_mix_NLTE_l0vTeff} we show the photometric observables
for the radial modes computed with two sources of the opacity tables and two chemical mixtures.
Here, we compared computations obtained with the OPAL $vs.$ OP tables
and the GN93 $vs.$ AGSS09 mixture. Again, the largest effects are for the p$_1$ mode
in both the amplitude ratios and phase differences.
The amplitude ratios computed with GN93 are smaller than those obtained with AGSS09,
because of relatively higher abundance of iron in AGSS09.
Also computations with the OP data give smaller values of $A_u/A_y$ comparing to those obtained with the OPAL data.
Moreover, the maximum of the OP amplitude ratio is shifted towards higher effective temperature.
This is connected with the location of the $Z-$bump layer, which occurs at slightly higher temperature
in the OP data. A usage of different opacity data changes the value of $\varphi_u-\varphi_y$ by about 0.2 rad at most.
The chemical mixture has rather a minor effect on phase differences.
We can see that also these input data affect the photometric observables of radial modes far more
than the atmospheric parameters (cf. Fig.\,8 and 11).

Effects of the opacity tables and chemical mixture for nonradial modes with $\ell=1$ and 2
are shown in Fig.\,\ref{ef_tab_mix_NLTE_l1vTeff} and Fig.\,\ref{ef_tab_mix_NLTE_l2vTeff}, respectively.
These data affect the $\ell=1$ modes very subtle and are far less important than the NLTE effects (cf. Fig.\,9).
In the case of the $\ell=2$ modes effects of opacities and mixture are comparable to the NLTE effects (cf. Fig.\,10).

As we could see, generally, in the case of radial modes effects of pulsational parameters
on photometric observables are much larger than effects of atmospheric parameters discussed in previous section.
The opposite is true for nonradial modes which are more sensitive to the NLTE effects,
in particular in the case of dipole modes.

\section{Conclusions}

We have presented a comprehensive overview of possible sources of uncertainties
in theoretical values of the photometric amplitudes and phases of early B-type pulsators.
These data are of particular importance because they serve as tools for mode identification
in the case of main sequence pulsators,
as well as contain information on stellar physics.

The uncertainties are embedded in stellar model atmospheres and nonadiabatic theory of stellar pulsation.
The atmospheric input consists of the flux derivatives over $T_{\rm eff}$ and $\log g$,
and limb darkening and its derivatives, in photometric passbands.
These quantities are sensitive to the atmospheric metallicity and microturbulent velocity, as well as
to the departure from the LTE approximation.
From pulsation computations, we get the nonadiabatic complex parameter, $f$, whose value depends
on chemical composition, opacities and subphotospheric convection if present.
If the $f$-parameter can be derived from observations then we get an extra seismic probe
by means of which we can test this input physics.

We begun with computations of tables with various quantities, needed to evaluate
the light variation in a given photometric band, for NLTE-TLUSTY model atmospheres.
These tables include data on the passband flux, flux derivatives over effective temperature and gravity,
and the non-linear limb darkening coefficients in 12 most popular passbands: $uvbyUBVRIJHK$.
All these data are public available and can be retrieved from the Wroc{\l}aw HELAS web page.

Then, we studied effects of these input parameters on the photometric observables, i.e. amplitude ratios
and phase differences. We considered the 10 $M_{\odot}$ main sequence models and the low degree modes with $\ell=0,1,2$.
In particular, the effect of NLTE model atmospheres was studied for the first time.
In the case of radial modes, this effect is comparable to an influence of the atmospheric metalicity and microturbulent velocity.
The photometric observables of nonradial modes appeared more sensitive to the departure from LTE.
Subsequently, we drew a parallel to effects of parameters related to pulsation computations.
These comparisons showed that the photometric observables of the radial modes
are by far more sensitive to the pulsation input. In turn, in the case of nonradial modes
the NLTE effect became more important, especially for the $\ell=1$ modes.

A complete seismic model should reproduce not only pulsational frequencies but also the observed values
of the photometric amplitude and phases, which can be translated into the empirical values of the $f$-parameter.
Our studies showed that such comprehensive approach should take into account all inaccuracies
in the photometric amplitudes and phases.

\begin{acknowledgements}
This work was supported by the HELAS EU Network, FP6, No. 026138
and for JDD by the Polish MNiSW grant N N203 379636.
\end{acknowledgements}

\end{document}